# Gauge-Programmable Mode Selection in Non-Hermitian Directed-Graph Networks

Wenwen Liu[1,2], Shuang Zhang[1,2,3,4,5§]

[1]*New Cornerstone Science Laboratory, Department of Physics, The University of Hong Kong, Hong Kong*
[2]*State Key Laboratory of Optical Quantum Materials, University of Hong Kong, Hong Kong 999077, China.*
[3]*Quantum Science Center of Guangdong-Hong Kong-Macao Great Bay Area, 3 Binlang Road, Shenzhen 518000, China.*
[4]*Materials Innovation Institute for Life Sciences and Energy (MILES), HKU-SIRI, Shenzhen 518000, China.*
[5]*Department of Electrical & Electronic Engineering, The University of Hong Kong, Hong Kong 999077, China.*

Programmable mode selection is essential for integrated photonic networks, yet selectively isolating individual modes without delicate balancing of gain and loss remains challenging. Here we introduce a gauge-engineering method in non-Hermitian directed-graph networks that support geometry-protected pure decay modes—eigenstates exhibiting smooth exponential amplitude decay along directed paths. In fully connected configurations, a single dominant mode emerges naturally, separated from the remaining modes by a large, tunable energy gap. By introducing synthetic gauge fields through phase-compensated non-reciprocal hopping, any desired pure decay mode can be promoted to the dominant position while its amplitude profile is preserved, which can be naturally interpreted by spectral graph theory. The approach further extends to simultaneous selection of paired modes in half-connected graphs and customizable multi-mode distributions in higher dimensions via orthogonal folding. Our method enables robust, loss/gain-free control over mode profiles, advancing applications in single-mode lasers, sensors, and quantum processing

## 1. Introduction

Non-Hermitian physics provides a versatile framework for describing open quantum and classical systems [1-4], manifesting complex eigenvalues [4, 5], exceptional points [6-9], and phenomena such as enhanced sensitivity [10, 11], directional transport [12, 13], and topological transitions [14, 15]. A central challenge is to selectively isolate and manipulate individual eigenmodes while suppressing competing modes [16-21]. Such control is crucial for maintaining coherence, stability, and functionality in a range of open-system platforms, including high-performance lasers and sensors [22-28].

Directed graphs have recently been identified as a class of non-Hermitian networks with engineered asymmetric connectivity and non-reciprocal hopping that support a complete basis of pure decay modes [13]. In these systems, all eigenstates exhibit smooth, non-oscillatory exponential decay profiles, with amplitudes decaying monotonically along directed paths—protected purely by the graph geometry without requiring loss/gain balancing or exceptional points. This feature arises under specific connectivity conditions, such as closed rings with alternating forward- and backward-directed chains or tailored graph topologies, distinguishing them from generic non-Hermitian systems prone to oscillatory skin effects.

In this work, we propose a gauge-engineering approach for flexible and precise mode selection within such directed-graph systems [29, 30]. In fully connected configurations meeting the required connectivity, a singular mode emerges with a large energy gap that can be tuned extensively by varying hopping ratios and site number. Specifically, a similarity transformation maps the system onto a circulant matrix, whose eigenmodes share the same exponential amplitude envelope with different phase windings. By introducing synthetic gauge fields through phase compensation in the hopping terms—without modifying loss or gain—we selectively promote any desired winding mode to the singular position with single tunning parameter while preserving its amplitude profile. Such framework extends to simultaneous double-mode selection in half-connected graphs, yielding paired modes separated from all others. A further extension to higher dimensions through orthogonal directional folding yields customizable multimode spectra over broad frequency ranges, with independent phase control along each dimension. Unlike previous methods based on exceptional points, parity–time symmetry, or structural defects [31-33], our approach provides robust control of mode profiles without gain or loss engineering; their protection arises entirely from the underlying graph geometry and gauge configuration.

## 2. Emergence of pure decay dominant mode

We start with directed-graph lattices featuring engineered asymmetric connectivity and non-reciprocal hopping. These lattices support a complete basis of pure decay modes characterized by purely imaginary eigenvalues and smooth, non-oscillatory amplitude profiles that decay exponentially along directed paths [13]. Figure 1(a) shows the configuration, in which the spheres denote lattice sites ( $m \in [4, N-2]$ for intermediate sites) and the yellow arrows indicate pairs of non-reciprocal hopping. The lattice Hamiltonian reads:

$$H = \sum_{i=1}^{N} \sum_{j=1, j\neq i}^{N} (t a_{j-i} c_i^{\dagger} c_j + a_{j-i} c_j^{\dagger} c_i) \quad (1)$$

where $a_{j-i} = 1$ or 0 is the binary coefficient and $t = \frac{t_r}{t_l}$ is the hopping ratio, with left- and right- $t_l$ and $t_r$ denoting the left- and right-directed hopping amplitudes, respectively. Onsite energies are neglected, as they are uniform across all nodes. When $a_{j-i} = a_{N-j+i}$, the system supports pure decay modes [13], with all eigenstates exhibiting non-oscillatory decay profiles, as shown in Fig. 1b. Of particular interest is the fully connected directed-graph (FCG) configuration in Fig. 1(a), for which every hopping coefficient is nonzero. In this case, one pure decay mode shifts upward along the imaginary axis (marked by the red dot in Fig. 1c with $N = 30$), whereas the remaining modes form a 'ground state' with lower imaginary energy. The resulting spectral separation produces a single dominant mode.

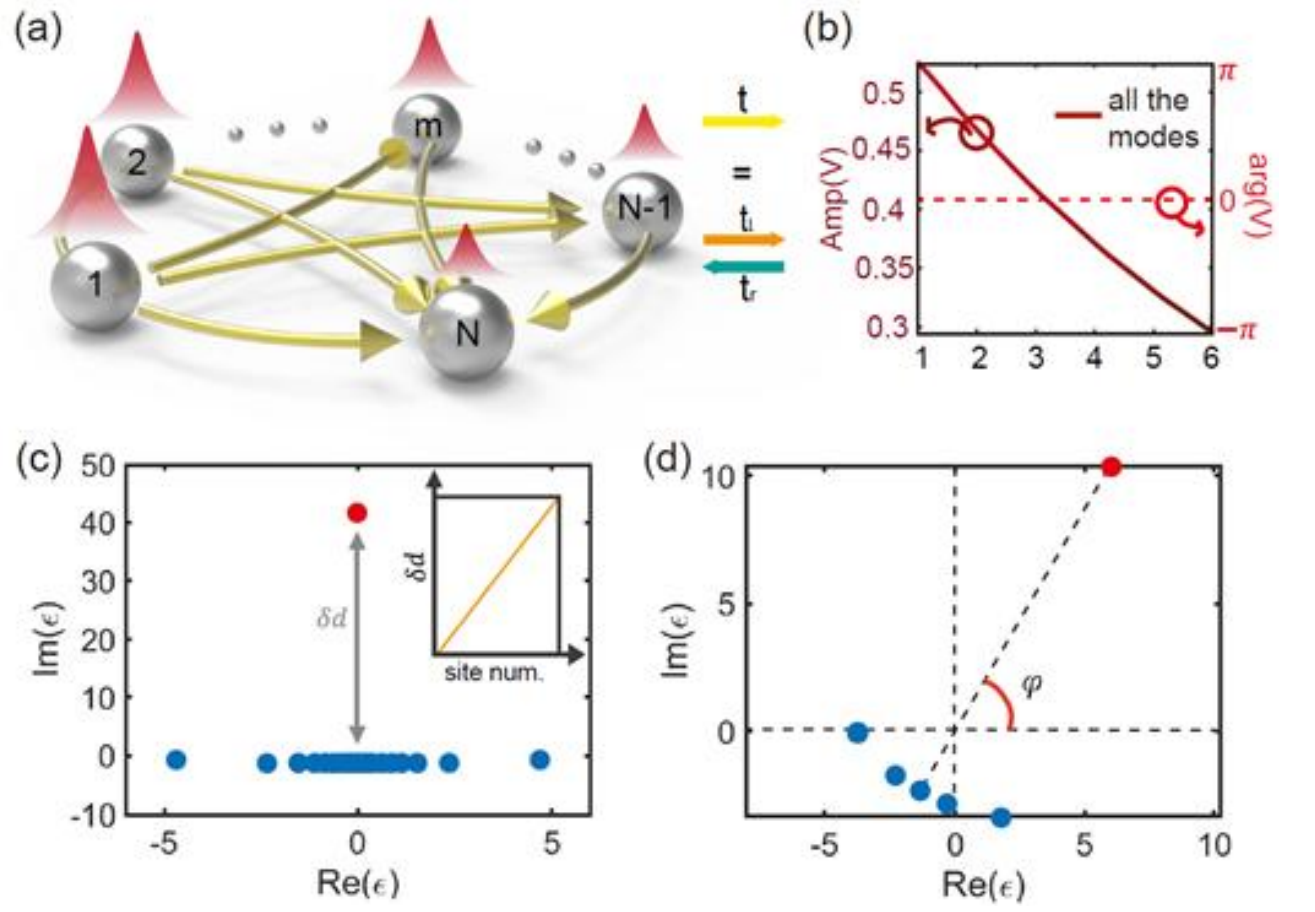

**Figure 1. Non-Hermitian directed graph that supports single-mode dominance.** (a) Schematic of the fully connected directed graph (FCG) with non-reciprocal hopping represented by yellow arrows which are equivalent to the left hopping $t_l$ and right hopping $t_r$ pairs. Sites are indexed 1 to *N*, with intermediate sites labelled *m*. The above pattern illustrates the pure decay mode which shows smooth exponential decay. (b) Amplitude distribution of all the modes (left) and phase distribution of singular mode (right). (c) Complex eigenenergy spectrum for $N = 30$, showing the singular mode (red dot) separated from the degenerate ground-state manifold (blue). The inset shows monotonic increase of the gap with increasing *N*. (d) Tunable spectrum rotation in the complex plane with complex hopping $t_l = 2e^{i\varphi}$ and $t_r = 1$, the singular behavior preserved.

To explain the emergence of this isolated mode, we solve the Hamiltonian using the ansatz $\psi_{mn} = r^{m-1} e^{i(n-1)(m-1)\theta}$ (with $r = t^{-1/N}$ and $\theta = 2\pi/N$ for simplification), which represents the $n_{th}$ eigen-state on the $m_{th}$ site. Focusing on the first site where all directional arrows pointing inward, the corresponding eigen-energy is

$$E_n = \sum_{q=1}^{N-1} a_q \, t^{\frac{N-q}{N}} e^{iq(n-1)\theta} \quad (2)$$

When all connectivity coefficients satisfy $a_q = 1$ (fully connected case), as in the fully connected case, the expression reduces to the geometric series $t(x + x^2 + \cdots + x^{N-1})$, where $x = re^{i(n-1)\theta}$. The sum evaluates to

$$E_n = t(-1 + \frac{1 - t^{-1}}{1 - re^{i\phi}}) \quad (3)$$

with $\phi = (n-1)\theta$. The maximum (least negative) value occurs when $e^{i\phi} = 1$ (i.e. $\phi = 2\pi m$ for integer $m$), which corresponds to $n = 1$ in this system. The phase distribution for this mode is shown by dashed line in Fig. 1b. As $r$ approaches 1 when $N > t$ (see more details in Supplemental Material), the remaining modes are well separated from the first, supporting the single pure decay mode.

For a fixed hopping ratio, the selectivity of the singular mode—quantified by the energy gap in the imaginary part of the eigen energies between the dominant mode and the degenerate manifold—improves with increasing system size *N*. As shown in the inset of Fig. 1c, the energy gap increases monotonically and becomes substantially larger with *N*. Furthermore, the position of the singular mode in the complex energy plane can be tuned arbitrarily by varying the complex hopping coefficients. For example, Fig. 1d demonstrates results with $t_l = 2e^{i\varphi}$ and $t_r = e^{i\varphi}$, where the entire spectrum is rotated by $\varphi$ while preserving the singular mode's dominance ($\varphi = \pi/3$ in this case). When all hopping coefficients are real, the single-mode separation occurs along the real energy axis (see more examples in Supplemental Material).

## 3. Graph interpretation and Gauge-engineered mode selection

Significantly, such system admits a natural spectral-graph interpretation. After subtracting the uniform onsite term, $H$ is a zero-diagonal complex weighted adjacency matrix. In the fully connected case, defining $D = diag(1, r, \ldots, r^{N-1})$ gives a circulant matrix with similarity transformation $C = D^{-1}HD$, or equivalently a convolution operator on the cyclic group $Z_N$(see supplemental materials for more details). The eigenvectors of C are the Fourier characters $\chi_n = \frac{1}{\sqrt{N}}\left(1, e^{i\phi_n}, \dots, e^{i(N-1)\phi_n}\right)^{\mathrm{T}}$, and the right eigenvectors of $H$ are $\psi_n = D\chi_n$. Thus, all pure-decay modes possess the same exponential amplitude envelope and are distinguished only by their phase winding as shown in Fig. 2(b). In the preceding analysis, the directed-graph geometry intrinsically selects the mode with phase winding $\phi = 0$ (colored red in Fig. 2(b)). We next show how to target any desired mode via gauge engineering.

The gauge is implemented by adding a compensating phase to each hopping term, as illustrated by the six-site example in Fig. 2(a). Different colors denote distinct non-reciprocal phase tunings, with positive phase accumulation along the forward (arrow) direction and negative accumulation in the backward direction. Note that the added phase follows the original hopping arrow direction. The phase for each hop is proportional to the site-number difference between connected sites, yielding distinct values (e.g. from $e^{i\theta}$ to $e^{i(N-1)\theta}$) differentiated by color. The resulting gauged Hamiltonian is

$$H_g = \begin{bmatrix} 0 & a_1 t e^{i\theta k} & \dots & a_{N-1} t e^{i(N-1)\theta k} \\ a_1 e^{-i\theta k} & 0 & a_1 t e^{i\theta k} & \dots \\ \dots & a_1 e^{-i\theta k} & 0 & a_1 t e^{i\theta k} \\ a_{N-1} t e^{-i(N-1)\theta k} & \dots & a_1 e^{-i\theta k} & 0 \end{bmatrix} \tag{4}$$

where $k \in [0, N-1]$ indexes the target mode. The eigenenergies become

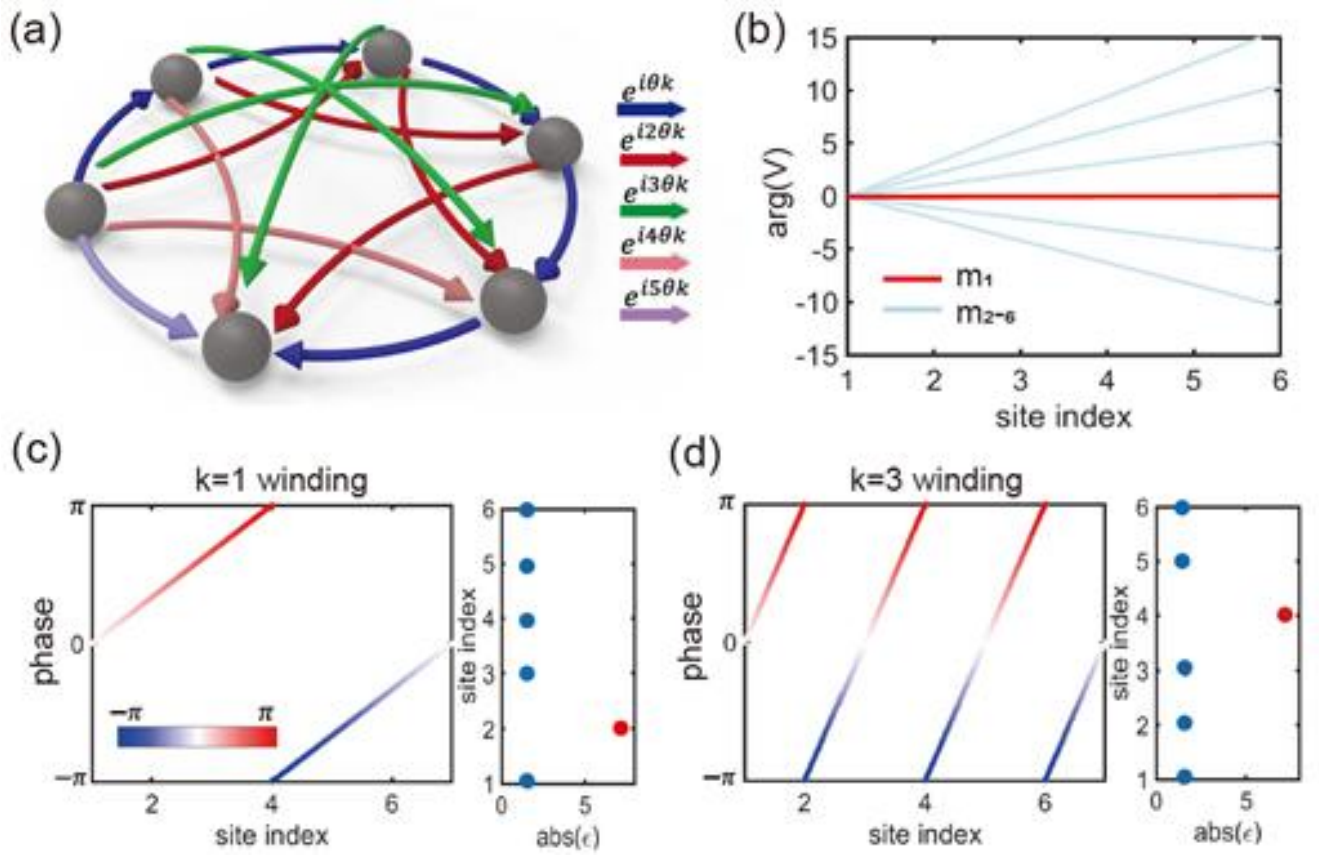

**Figure 2. Gauge-engineered single-mode selection**. (a) 6-site example of synthetic gauge addition; colors indicate phase compensation (positive along arrow direction, negative opposite). (b) Phase distributions of all pure decay modes in the ungauged case (same amplitude profile). (c, d) The phase profile of selected modes after gauge application: for $k = 1$ (c) and $k = 3$ (d), the 2nd and 4th modes become singular (left panels), with corresponding phase distributions shown different winding number from $-\pi$ to $\pi$. Amplitude profiles remain unchanged (identical to Fig. 1b).

$$E_n = t\left(-1 + \frac{1 - r^N}{1 - re^{i(\phi - \theta k)}}\right) \tag{5}$$

Figure 2(c-d) presents the results of single-mode selection under different gauge configurations. For $k = 1$ ($k = 3$), the 2nd (4th) mode is clearly selected, with its phase distribution exhibits $2\pi$ and $6\pi$ winding. Notably, all selected modes retain the same amplitude profile as in the ungauged case (Fig. 1(b)). This result also has a graph-theoretic interpretation. For the discrete gauge parameter k, the phase compensation is equivalent to a diagonal unitary similarity transformation $H^k = U_k^{-1} H U_k$, where $U_k = diag\left(1, e^{-ik\theta}, \dots, e^{-i(N-1)k\theta}\right)$. Therefore $E_n^k = E_{n-k}^0$, with indices understood modulo N. The eigenvalue set and the spectral gap are consequently unchanged. Thus, the gauge acts as a Fourier shift which cyclically reassigns the pre-existing isolated extremal branch to the selected phase-winding mode while leaving its spatial amplitude profile invariant.

Moreover, since such phase-only gauge transformation does not change the spectrum, the complete set of modal thresholds including the lowest threshold is preserved. Meanwhile, only the phase-winding eigenvector that occupies the minimum-threshold spectral branch changes. The gauge transformation therefore switches between winding configurations without altering either the linear threshold hierarchy or the spatial amplitude distribution.

## 4. Double mode selection

The fully connected graph is replaced by a half-connected state (HCS), in which each site is connected only to sites two steps away (i.e., one intermediate site is skipped). This restricts the hopping to odd- and even-numbered sites only, yielding a bipartite graph structure with no intra-sublattice connections, as illustrated in Fig. 3(a). The Hamiltonian has the same form as in the FCG case, except that $a_i = 0$, for $i \in 2Z$ (or equivalently, only odd-indexed connections are active). Consequently, the eigenenergy at the reference site therefore reduces to

$$E_n = \sum_{p=1}^{N/2} a_{2p-1}\, t^{\frac{N-(2p-1)}{N}} e^{i(2p-1)(n-1)\theta} \tag{6}$$

which forms a geometric series $t\frac{x(1-x^N)}{1-x^2} = \frac{1-t}{x-\frac{1}{x}}$ with $x = re^{i(n-1)\theta}$. The maximum value occurs when $e^{i2\phi} = 1$ ($\phi = (n-1)\theta$), supporting exactly two modes: $n_1 = 1$ and $n_2 = N/2 + 1$. These modes are clearly separated from the others, as shown by the red points in Fig. 3(b) for $t_l = 1.3i$, $t_r = 1i$ (marked by red dots). They share the same absolute eigenenergy magnitude but are well separated in the complex plane (Fig. 3(c)).

Closer inspection reveals that the two modes share the same amplitude distribution as in the single-mode case but exhibit distinct phase profiles: one in-phase and the other out-of-phase across the two sublattices, as illustrated for the six-site system in Fig. 3(d). Gauge can be introduced in a similar manner as in the single-mode case, as shown in the inset of Fig. 3(d). Varying the gauge parameter $k$ selects any desired pair while retaining exactly two isolated modes. The corresponding results are shown in Supplemental material. This behavior follows from the eigenenergy expression: the gauge shifts the phase condition so that two values of $\phi$ satisfy the maximization criterion. As with the single-mode case, these paired modes can be positioned anywhere in the complex energy plane by tuning the hopping coefficients $t_l$ and $t_r$.

These two isolated modes, however, cannot both be dominant under the preceding parameter choice because they cannot simultaneously attain the minimum loss. Two dominant modes can be obtained by choosing a negative value of $t$. The associated phase compensation organizes all states into complex-conjugate pairs, namely, i.e., $t^{-1/N}e^{i\theta}$ & $t^{-1/N}$ or $t^{-1/N}e^{i(\pi-\theta)}$ & $t^{-1/N}e^{i\pi}$ . Consequently, we obtain $E_{n_1} = E^*_{n_2}$ and $E_{n_3} = E^*_{n_4}$, where $n_1 = 1$, $n_2 = 2$, $n_3 = N/2 + 1$ and $n_4 = N/2$ . The case $t = -2$, shown in Supplemental Material, contains four isolated states: two shift upward and two shift downward along the imaginary axis, so that two dominant modes emerge simultaneously. Adding a gauge field then permits flexible selection of any desired dominant pairs.

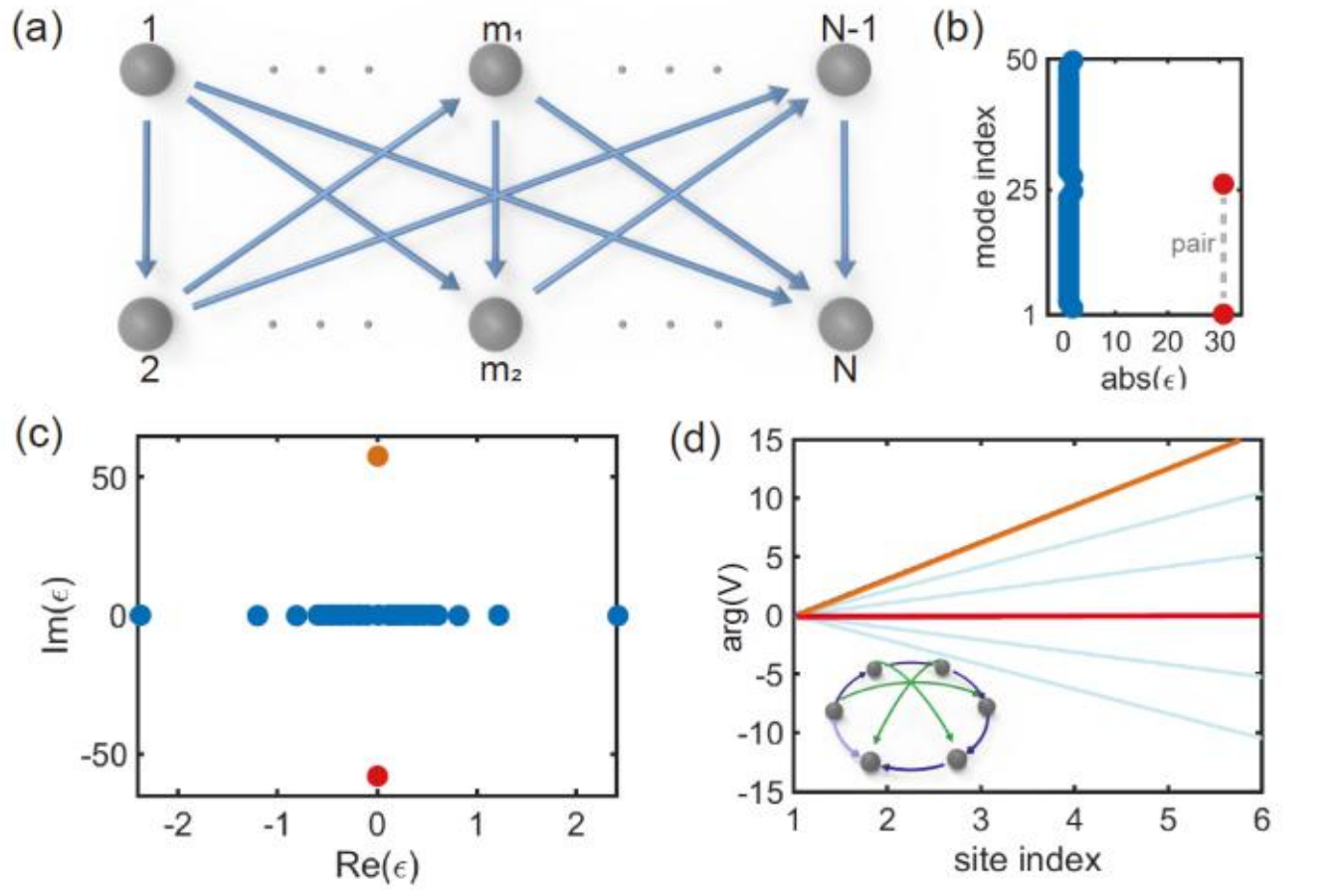

FIG. 3. **Realization of double mode selection.** (a) Graph configuration with hopping only between odd and even sites (skipping one intermediate site), forming a bipartite structure with directed non-reciprocal coupling. (b) Absolute eigenenergy values for $N = 20$ ($t_l = 2$, $t_r = 1$); selected modes (bright/red) well separated from others (blue). (c) Energy spectrum with two dominant modes pair highlighted in red and orange. (d) Phase profiles of the two modes (in-phase and out-of-phase across sublattices). The inset shows 6-site example of gauge addition.

## 5. Mode selection for multiple frequency

We next generalize the theory to select multiple modes distributed over a broad frequency range by through dimensional extension. The construction involves two steps. First, we show that the eigenenergy obey a nesting rule: the spectrum in one dimension extends or folds over that in another dimension, and the process can be continued in additional dimensions. Second, by choosing and combining suitable eigenspectra for the individual dimensions, we can construct the desired spectral distributions.

Obtaining analytical eigen-solutions requires that the boundary conditions and bulk equations be satisfied independently for each orthogonal direction. Meanwhile, the Hamiltonian contains only principal-axis hopping and can be written as a Kronecker sum. The system eigenmodes are therefore direct products of the eigenmodes obtained independently in each dimension [13]. For the two-dimensional system shown schematically in Fig. 4(a), any eigenmode at site $(m_x, m_y)$ can be expressed as

$$\psi_{m_x,m_y,n_x,n_y} = \psi_{m_x n_x}\psi_{m_y n_y} \tag{7}$$

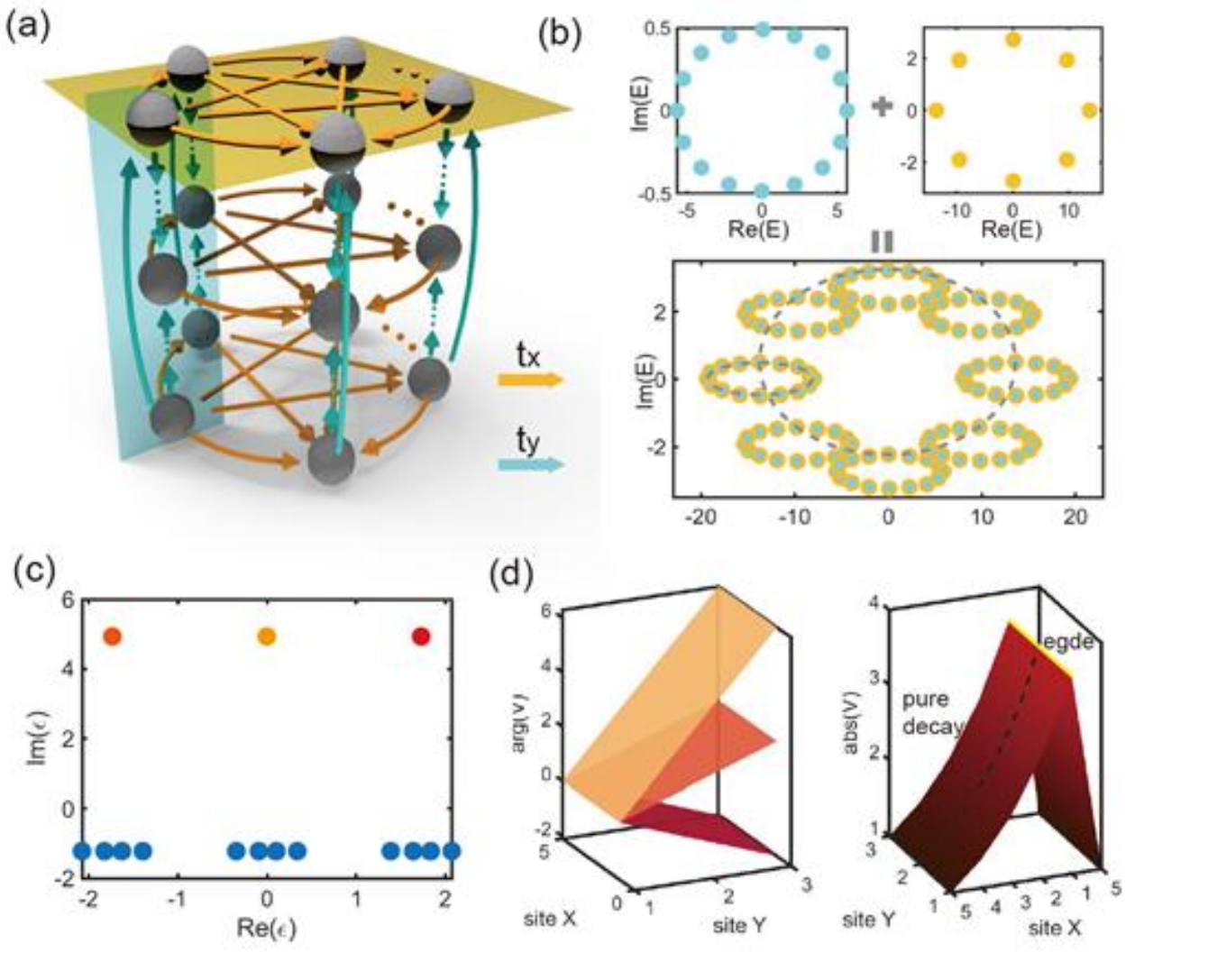

**Figure 4. Extension to multi-mode selection via dimension folding.** (a) 2D graph configuration with independent non-reciprocal hopping in x and y directions. (b) General illustration of Energy folding. (c) Resulting spectrum with three selected modes with the same imaginary energy but differing in real frequency. (d) amplitude and phase distributions for the three selected modes. The phases are distinguished by color that corresponds to the dominant modes in (c). The amplitudes are identical across modes with pure decay profile and edge state along y direction.

where $\psi_{m_x n_x} = t_x^{(m_x-1)/N_x} e^{i(n_x-1)(m_x-1)\theta_x}$ and $\psi_{m_y,n_y} = t_y^{(m_y-1)/N_y} e^{i(n_y-1)(m_y-1)\theta_y}$ with $\theta_x\,(\theta_y) = 2\pi/N_x (2\pi/N_y)$, and the corresponding eigenenergy is additive: $E_n = E_{nx} + E_{ny}$. We focus on the corner site where all hoppings point inward. In this case, $E_{n\alpha}$ ($\alpha = x, y$) can be obtained by summing over incoming hops in that dimension as in the 1D case. According to the dimensional nesting, the total number of distinct eigenenergy equals the least common multiple of $N_x$ and $N_y$. Consequently, each spectrum from one dimension repeats the full set of values from the other, e.g., each $E_{nx}$ appears $N_y$ times paired with every possible $E_{ny}$ as shown in Fig. 4(b).

To achieve multiple-mode selection over a broad real-frequency region, we extend the single-mode configuration (supporting a dominant mode on the imaginary axis) with pure-phase non-reciprocal hopping in the additional dimension. For $t_{l_x} = 1.5i$ and $t_{r_x} = 1i$ along $x$, and pure-phase hoppings $t_{l_y} = e^{i\chi_1}$ and $t_{r_y} = e^{i\chi_2}$ along $y$, three selected modes span a range of real frequencies [Fig. 4(c)]. Their phase and amplitude distributions are shown in Fig. 4(d). The amplitude profiles remain uniform and smoothly decay across all selected modes, while the phases vary according to the dimensional structure. Applying an appropriate gauge along a chosen dimension similarly selects distinct phase profiles in that direction without affecting the other dimensions (see more details in Supplemental Material). This construction therefore provides a general framework for controllable multimode

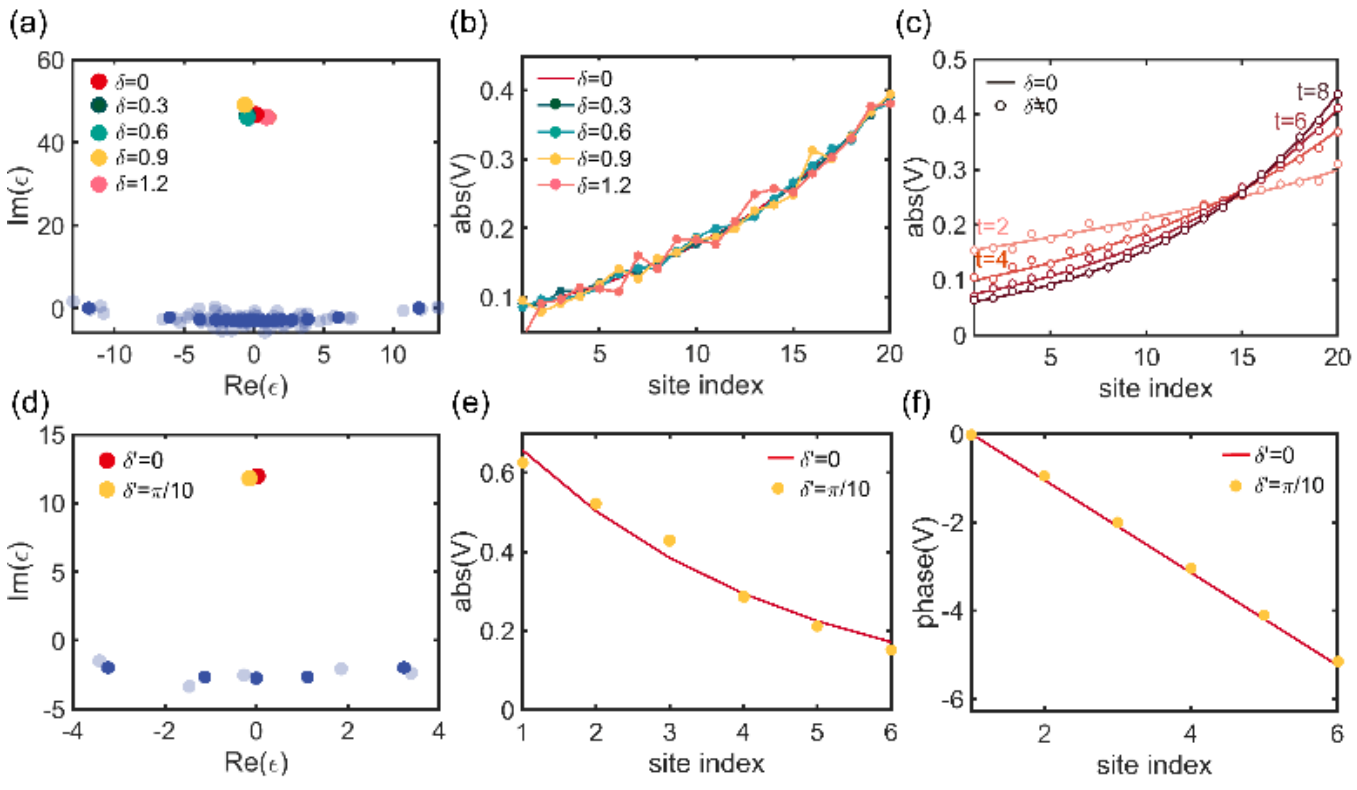

**Figure 5. Robustness against disorder**. (a) Complex eigenvalue spectra and (b) selected-mode amplitude profiles for a 20-site system with $t$=5 and different disorder strengths. (c) Influence of the hopping ratio t on the selected-mode profile at fixed disorder. (d–f) Complex spectrum, amplitude profile, and phase profile of the gauge-selected mode in a 6-site system with phase disorder $\delta'$=π/10. Solid curves denote the disorder-free results, whereas symbols denote representative disorder realizations.

selection and distribution across real frequencies.

## 6. Robustness and experimental feasibility

Experimental implementations inevitably contain fabrication imperfections and parameter fluctuations. Stability and robustness against disorder are therefore important criteria for evaluating the practical performance of the proposed mode-selection mechanism. In the following, we investigate the influence of disorder by adding independent random perturbations to all hopping and onsite terms.

The perturbation is added with $p = \delta\,\mathrm{rand}[-1,1]$, where δ represent the disorder strength. For the 20-site system with $t = 5$, the selected eigenvalue remains clearly separated from the competing modes throughout the investigated range $0 \le \delta \le 1.2$ [Fig. 5(a)]. Its eigenvector also retains an approximately exponential decay envelope, although local fluctuations increase with disorder. Such deformation can be mitigated by increasing the relative hopping ratio t, as shown in Fig. 5(c). The solid lines represent the original decay profiles, whereas the open circles represent the profiles in the presence of disorder. The corresponding profile residuals for $t = 2, 4, 6, 8$ are $[3.1, 1.1, 0.9, 0.3] \times 10^{-3}$, respectively. Therefore, for a fixed disorder strength, the selected-mode profile is less affected when a larger hopping ratio is used.

We further introduce gauge-phase disorder according to $g_{ij} = \mathrm{e}^{i\left(k\theta_{ij} + \delta' \mathrm{rand}(-1,1)\right)}$. For the six-site example with $\delta' = \pi/10$, the target mode remains spectrally isolated, while its amplitude and phase profiles remain close to their disorder-free values [Fig. 5(d–f)]. These results demonstrate that the proposed mode-selection mechanism remains effective under moderate hopping, onsite, and phase disorder.

Scalability is another important concern in experiment. The fully connected construction requires $N(N-1)$ directed hopping terms and therefore entails hardware cost. Although the link phases follow the structured relation $\phi_{ij} = (j-i)\alpha$ and are determined by a single gauge parameter, each physical link must still realize the prescribed hopping. This establishes an explicit trade-off between selectivity and implementation resources. Nevertheless, systems with appropriate N, such as the six-site examples considered here can realize great performance requiring only a moderate number of couplers and are compatible with currently available integrated-resonator, waveguide-network, or coupled cavity arrays [13, 34-36]. In these systems, each site corresponds to one resonator or waveguide mode, while each directed edge corresponds to an engineered complex hopping coefficient with a prescribed amplitude asymmetry and phase. The non-reciprocal hopping ratio determines the exponential envelope of the pure decay modes, whereas link phases supplied by thermo-optic, electro-optic, or dynamically modulated couplers provide the synthetic gauge field used to select the target mode. Larger implementations may instead benefit from shared-bus, multiplexed, dynamically modulated platforms or synthetic-dimension coupling architectures.

Overall, the system remains robust under substantial perturbations. Connected networks with approximately five or six sites are already within the reach of current microwave-resonator circuits platforms while more complex configurations may become feasible through further advances in device fabrication, coupling control, and scalable modulation technologies. Moreover, under the convention adopted here, the gauge-selected target state always has the lowest lasing threshold when gain is introduced, therefore could lase before the remaining modes. The link-phase control can consequently be employed to determine the first lasing mode, without requiring a redesign of the underlying gain–loss distribution. This establishes a practical route toward reconfigurable single- and multimode photonic devices whose modal response can be programmed directly through the synthetic gauge field.

### 7. Conclusion

In summary, we have introduced a gauge-engineering framework for programmable mode selection in non-Hermitian directed-graph networks. Such systems can be mapped to circulant matrices and naturally support a complete set of geometry-protected pure decay modes. The modes share a smooth, non-oscillatory exponential amplitude envelope along directed paths and differ in their phase winding. By strategically incorporating synthetic gauge fields through phase-compensated non-reciprocal hopping, we demonstrate that any desired pure decay mode, or pair of modes, can be selectively promoted to a dominant position while faithfully preserving its characteristic amplitude profile. A single tunable gauge parameter provides precise control, eliminating the need for delicate gain-loss balancing or operation near exceptional points.

The approach extends naturally to half-connected bipartite graphs for simultaneous double-mode selection and to higher-dimensional architectures via orthogonal directional folding, enabling customizable multi-mode spectra distributed across broad frequency ranges. In all cases, the selected modes remain robustly protected by the underlying graph geometry and gauge configuration. numerical results and analytical derivations confirm the large, tunable energy gaps and the versatility of this method across different hopping configurations and system sizes.

The framework can be further enriched by incorporating dynamic gauge modulation, nonlinear interactions, or topological features, thereby opening new avenues for non-Hermitian wave engineering and active photonic devices. We anticipate that this work will stimulate both theoretical explorations and experimental realizations of geometry- and gauge-protected functionalities in non-Hermitian physics.

§To whom correspondence should be addressed
shuzhang@hku.hk